\documentclass[conference]{IEEEtran}

\usepackage{amsmath} % assumes amsmath package installed
\usepackage{amssymb}  % assumes amsmath package installed
\usepackage[utf8]{inputenc}
\usepackage[amsmath,thmmarks,hyperref]{ntheorem} % nicer theorems
\usepackage{bbm}               % nicer bbm fonts
\usepackage{mathtools}         % some nice tricks for stackrel
\usepackage{stmaryrd}          % more brackets
\usepackage[final]{graphicx}   % to include pdf pictures
\usepackage{paralist}          % better enumerate lists
\usepackage{aliascnt}          % TEST alias counter for ntheorem environments
\usepackage{subcaption} %for subfigures
\usepackage{mathrsfs}

\usepackage{textcomp}
\graphicspath{{pics/}}

\newcommand\MTkillspecial[1]{% helper macro
\bgroup
\catcode`\&=9
\let\\\relax%
\scantokens{#1}%
\egroup
}

\renewcommand{\mathbb}[1]{\mathbbm{#1}}

\renewcommand{\theenumi}{\roman{enumi}}
\newcommand{\refitem}[1] {\textit{\ref{#1}.)}}

\theoremheaderfont{\normalfont\itshape}
\theorembodyfont{\normalfont}
\newtheorem{lemma}{Lemma}

\newaliascnt{proposition}{lemma}
\newtheorem{proposition}[proposition]{Proposition}
\aliascntresetthe{proposition}

\newaliascnt{thm}{lemma}

\aliascntresetthe{thm}

\newaliascnt{corollary}{lemma}

\aliascntresetthe{corollary}

\newaliascnt{definition}{lemma}
\newtheorem{definition}[definition]{Definition}
\aliascntresetthe{definition}

\newaliascnt{claim}{lemma}

\aliascntresetthe{claim}

\theorembodyfont{\normalfont}

\newaliascnt{example}{lemma}

\aliascntresetthe{example}

\newaliascnt{remark}{lemma}
\newtheorem{remark}[remark]{Remark}
\aliascntresetthe{remark}

\newaliascnt{question}{lemma}

\aliascntresetthe{question}

\newaliascnt{conjecture}{lemma}

\aliascntresetthe{conjecture}

\theorembodyfont{\normalfont}
\theoremnumbering{Roman}

\def\theorem@checkbold{}

\newenvironment{propositionlist}{\begin{compactenum}[\itshape i.)]}{\end{compactenum}}

\allowdisplaybreaks

\let\originalleft\left
\let\originalright\right
\renewcommand{\left}{\mathopen{}\mathclose\bgroup\originalleft}
\renewcommand{\right}{\aftergroup\egroup\originalright}

\newcommand{\I}              {\mathrm{i}}
\newcommand{\E}              {\mathrm{e}}
\newcommand{\algebra}[1]      {\mathscr{#1}}
\newcommand{\basis}[1]       {\mathit{#1}}

\newcommand{\tensor}[1][{}]           {\mathbin{\otimes_{\scriptscriptstyle{#1}}}}
\DeclarePairedDelimiter\ideal\langle\rangle
\reDeclarePairedDelimiterInnerWrapper\ideal{star}{
\mathopen{#1\vphantom{\MTkillspecial{#2}}\kern-\nulldelimiterspace\right.}
#2
\mathclose{\left.\kern-\nulldelimiterspace\vphantom{\MTkillspecial{#2}}#3}}

\DeclarePairedDelimiter{\dualpair}{(}{)}

\newcommand{\colvec}[3]{\tiny
    \begin{pmatrix}#1\\#2\\#3\end{pmatrix}
}

\newcommand{\DAT}{\mathsf{DCT}}

\begin{document}
\IEEEoverridecommandlockouts

\title{Fast cosine transform for FCC lattices
  }

\author{\IEEEauthorblockN{Bastian Seifert}
  \IEEEauthorblockA{\textit{Faculty of Engineering Sciences} \\
    \textit{Ansbach University of Applied Sciences}\\
    Ansbach, Germany \\
    bastian.seifert@hs-ansbach.de}
  \and
  \IEEEauthorblockN{ Knut Hüper}
  \IEEEauthorblockA{\textit{Institute of Mathematics} \\
    \textit{University of Würzburg}\\
    Würzburg, Germany \\    
    hueper@mathematik.uni-wuerzburg.de}
  \and \IEEEauthorblockN{Christian Uhl}
  \IEEEauthorblockA{\textit{Faculty of Engineering Sciences} \\
    \textit{Ansbach University of Applied Sciences}\\
    Ansbach, Germany \\
    christian.uhl@hs-ansbach.de} \\
  \thanks{This work is partially supported by the European Regional
    Development Fund (ERDF)}
} 

\IEEEpubid{\makebox[\columnwidth]{978-1-5386-5346-3/18/\$31.00~\copyright~2018
    IEEE \hfill} 
\hspace{\columnsep}\makebox[\columnwidth]{ } }
\maketitle
%\thispagestyle{empty}
%\pagestyle{empty}

%%%%%%%%%%%%%%%%%%%%%%%%%%%%%%%%%%%%%%%%%%%%%%%%%%%%%%%%%%%%%%%%%%%%%%%%%%%%%%%%
\begin{abstract}    
    Voxel representation and processing is an important
    issue in a broad spectrum of applications. E.g., 3D imaging in
    biomedical engineering applications, video game development and
    volumetric displays are often based on data representation by
    voxels. By replacing the standard sampling lattice with a
    face-centered lattice one can obtain the same sampling density
    with less sampling points and reduce aliasing error, as well. We
    introduce an analog of the discrete cosine transform for the
    face-centered lattice relying on multivariate Chebyshev
    polynomials. A fast algorithm for this transform is deduced based
    on algebraic signal processing theory and the rich geometry of the
    special unitary Lie group of degree four.
\end{abstract}


%%%%%%%%%%%%%%%%%%%%%%%%%%%%%%%%%%%%%%%%%%%%%%%%%%%%%%%%%%%%%%%%%%%%%%%%%%%%%%%%
\section{Introduction}
\label{sec:Introduction}%

The approximation of real world objects by voxel data and fast
processing of these 3D-samples is of great interest in a broad field
of applications. For example, in biomedical engineering one is
interested in fast processing of volumetric data in a variety of
3D-imaging procedures, like computed tomography, magnetic resonance
imaging and positron emission tomography to name just a few. In
computer games engineering there is an ongoing trend to rely graphics
on voxels instead of polygons. Another field of voxel processing is
volumetric displays requiring a large amount of bandwith for
refreshing. Hence, achieving the same sampling density by less data
would be highly desirable. Besides faster processing, reducing
aliasing errors and jitter noise by alternative data sampling
techniques is a motivation for the presented study. A method to
achieve this, is to change the sampling lattice from the standard
rectangular lattice, also known as cartesian cube (CC) lattice, to
either the face-centered cubic (FCC) lattice or the body-centered
cubic (BCC) lattice. The FCC lattice is associated to a densest sphere
packing, while the BCC lattice is interrelated to a sphere covering
\cite{Conway.Sloane:1999}. These lattices are in duality to each
other, i.e. the frequency data of data sampled on one lattice will sit
on the other lattice and vice versa. Both lattices have advantageous
sampling properties when compared to the CC lattice, each in its own
region of sampling frequency
\cite{Vad.Csebfalvi.Rautek.Groeller:2014,Kuensch.Agrell.Hamprecht:2005}.
For example, one could achieve the same sampling density on a FCC
lattice compared to a CC lattice with $29$ \% less sampling points.
Recently, tools for interpolation on these lattices were published,
e.g. the Voronoi splines~\cite{Mirzargar.Entezari:2010}. Furthermore,
it was shown that human beings recognize the sampled object at least
as good as if the data was sampled on a CC lattice
\cite{Meng.Entezari.Smith.Moeller.Weiskopf.Kirkpatrick:2011}. The
superiority of sampling on non-standard lattices has been shown in
medical applications \cite{Saranathan.Ramanan.Gulati.Venkatesan:2007},
as well.

Even though there are classical abstract sampling and reconstruction
theorems on arbitrary lattices \cite{Petersen.Middleton:1962} and
methodologies for decomposing general lattices into cartesian
sublattices for computing Fourier transforms
\cite{Mersereau.Speake:1981}, FFTs on FCC and BCC lattices have been
elaborated only recently \cite{Zheng.Gu:2014}. One big disadvantage of
these transforms is that they implicitly assume \emph{directed}
lattices, while 3D images are space-dependent objects, and hence it is
more natural to model them on \emph{undirected} lattices. Using the
theory of algebraic signal processing \cite{Pueschel.Moura:2008a} it
is easy to realize that 1D undirected lattices are connected to
discrete \emph{cosine} transforms, which are based on Chebyshev
polynomials \cite{Pueschel.Moura:2003a}, and deduce the corresponding
FFT-like algorithms \cite{Pueschel.Moura:2008c}. Relying on this
concept, an analog of the discrete cosine transform on the hexagonal
lattice, based on two variable Chebyshev polynomials, was derived
together with its fast algorithm \cite{Pueschel.Roetteler:2008}.
Recently, discrete cosine transforms on hexagonal lattices have gained
attention as feature generators for artificial neural networks in face
detection \cite{Azam.Anjum.Javed:2010}.

The multivariate Chebyshev polynomials \cite{Hoffman.Withers:1988} are
now classical, but have found applications only in the last years.
Apart from the applications in algebraic signal processing they are
applied in the discretization of partial differential equations in
\cite{Ryland.Munthe-Kaas:2011,Munthe-Kaas.Nome.Ryland:2012}, for the
derivation of cubature formulas in
\cite{Li.Xu:2010,Moody.Patera:2011,Hrivnak.Motlochova.Patera:2016} and
developing discrete transforms in \cite{Atoyan.Patera:2007}.

In this paper we are studying the derivation of an analog of the
discrete cosine transform on the FCC lattice and the corresponding
fast algorithm. Our derivation is based on algebraic signal processing
theory, which is recalled in
section~\ref{sec:AlgebraicSignalProcessing}, and Chebyshev polynomials
of the first kind in three variables, whose construction is shown in
section~\ref{sec:LieGroupsAndChebyshev}. The nice properties of the
Chebyshev polynomials in three variables are due to the connection to
the rich geometry of the Lie group $\mathsf{SU}(4)$, which is briefly
sketched. We derive the fast algorithm in
section~\ref{sec:CooleyTukeySU4} and apply the transform to an
artifical data set in section~\ref{sec:Application}.

%%%%%%%%%%%%%%%%%%%%%%%%%%%%%%%%%%%%%%%%%%%%%%%%%%%%%%%%%%%%%%%%%%%%%%%%%%%%%%%%
\section{Algebraic signal processing}
\label{sec:AlgebraicSignalProcessing}%

Algebraic signal processing theory~\cite{Pueschel.Moura:2008a} gives a
unified notion for the central concepts of linear signal processing.
The basic objects in this theory are \emph{signal models}, which
consist of triples $(\algebra{A}, M, \Phi)$, where $\algebra{A}$ is
the filter space, $M$ the signal space, and
$\Phi \colon \mathbb{C}^N \to M$ is a bijective map, called
$z$-transform. Typically, the filter and the signal space are choosen
to be equal as polynomials in multiple variables. The multiplication
of the polynomials is defined modulo some subset, termed ideal, $I$ of
polynomials, i.e.
$\algebra{A} = M = \mathbb{C}[x_1,\dots,x_n] \big/ I$. If we choose a
Gröbner basis~\cite{Adams.Loustaunau:1994} for the ideal this results
in multiplication modulo the polynomials in that Gröbner basis.
Choosing a basis in $M$ gives rise to the $z$-transform $\Phi$ by
mapping each vector entry to a coefficient of the basis elements of
$M$. For example choosing
$\algebra{A} = M = \mathbb{C}[z^{-1}] \big/ \ideal{z^{-n} - 1}$ with
basis $\{1, z^{-1}, z^{-2}, \dots, z^{-n+1}\}$ yields the well-known
$z$-transform from discrete, finite-time signal processing
\begin{equation}
    \label{eq:DiscreteFiniteTimeModel}
    \Phi \colon
    (s_0,  s_1,  \dots, s_{n-1})
    \mapsto
    s_0 + s_1 z^{-1} + \dots + s_{n-1} z^{-n+1}. 
\end{equation}
The zeros of $z^{-n} - 1$ are precisely the discrete frequencies
$\omega_n^k$, with $\omega_n$ being an $n$th root of unity. If the
zeros $\alpha_i$ of the ideal $I$ are distinct, we have, by the
Chinese remainder theorem, a decomposition of the signal space into
one-dimensional subspaces
\begin{equation}
    \label{eq:ChineseRemainderDecomposition}
    \begin{split}
        &\mathbb{C}[x_1,\dots,x_n] \big/ I \\
        &\cong \bigoplus_i \mathbb{C}[x_1,\dots,x_n] \big/ \ideal{x_1 -
          \alpha_{i,1}, \dots, x_n - \alpha_{i,n}}.
    \end{split}
\end{equation}
A matrix realizing this decomposition is called \emph{Fourier
  transform} of the signal model. In the case of the discrete, finite
time signal model this is precisely the discrete Fourier transform.

There are distinguished elements of the filter space which generate
the whole space. In the context of the algebraic signal processing
theory, they are called \emph{shifts}. In the example
$\mathbb{C}[z^{-1}] \big/ \ideal{z^{-n} - 1}$ the generator would be
$z^{-1}$. The generators allow for visualization of the signal model
by multiplying each basis element with the generators and drawing
arrows to the basis elements appearing in the result. 

%%% Local Variables:
%%% mode: latex
%%% TeX-master: "ControloPaper"
%%% End:

%%%%%%%%%%%%%%%%%%%%%%%%%%%%%%%%%%%%%%%%%%%%%%%%%%%%%%%%%%%%%%%%%%%%%%%%%%%%%%%%
\section{Permutations and generalized Chebyshev
  polynomials}
\label{sec:LieGroupsAndChebyshev}%

We briefly discuss the construction of multivariate Chebyshev
polynomials via permutation groups. The construction yields a basis of
the space of multivariate polynomials. This basis has useful properties,
which can be exploited to develop FFT-like algorithms.
We start by recalling the definition of the classical Chebyshev
polynomials of the first kind as
\begin{equation}
    \label{eq:UnivariateChebyshevPolynomials}%
    T_k(x)
    = T_k(\cos \theta)
    = \cos n \theta
    = \tfrac{1}{2} \left( \E^{ 2 \pi \I n \theta} + \E^{-2 \pi \I n \theta}
    \right) 
\end{equation}
for $\theta \in (0,1)$ and $x = \cos(\theta)$. The appearing numbers
$1$ and $-1$ can be interpreted as $1 \times 1$ matrices representing
the permutations of two elements. The number $2$ is then the number of
all such permutations, i.e. we average over all permutations. For a
generalization of Chebyshev polynomials in one variable to polynomials
in multiple variables, we replace the permutations of two elements
by another permutation group and represent it as matrices. Then we
multiply these matrices with vectors, exponentiate and average over
them. More formally, let
$\dualpair{k, \theta} := \exp(2 \pi \I k^\top \theta)$ for
$k \in \mathbb{Z}^d, \theta \in \mathbb{R}^d$, and denote by
$\dualpair{k, \theta}_s := \tfrac{1}{|W|} \sum_{w \in W} \dualpair{k,
  w \theta}$, termed $W$-symmetrization or generalized
cosine~\cite{Hoffman.Withers:1988}, for a permutation group $W$. The
definition of multivariate Chebyshev polynomials is then
straightforward:
\begin{definition}
    \label{definition:MultivariateChebyshevPolynomials}%
    Let $W$ be a permutation group. For each $k \in \mathbb{Z}^d$
    define the corresponding multivariate Chebyshev polynomials of the
    first kind as
    \begin{equation}
        \label{eq:MultivariateChebyshevPolynomials}
        T_k(x_1, \dots, x_d)
        := (k, \theta)_s
    \end{equation}
    with $x_j(\theta) := (\basis{e}_j, \theta)_s$ being a change of
    variables for $j = 1, \dots, d$, here
    $\basis{e}_j \in \mathbb{Z}^d$ denote the standard basis vectors.
\end{definition}
We list some of the properties multivariate Chebyshev polynomials
obey. 
\begin{proposition}
    \label{proposition:PropertiesOfMultivariateChebyshevPolynomials}%
    For the multivariate Chebyshev polynomials associated to the permutation
    group $W$ one has
    \begin{enumerate}
        \item \label{item:OneIsChebyshevPoly} $T_0(x_1,\dots, x_d) =
        1$,
        \item \label{item:MonomialsAreChebyshevPoly}
        $T_{\basis{e}_j}(x_1, \dots, x_d) = x_j$,
        \item \label{item:ChebyshevInvariantUnderWeylAction} $T_k =
        T_{w^T k}$ for all $w \in W$,
        \item \label{item:MotherRecurrenceRelation} the
        recurrence relation
        \begin{equation}
            \label{eq:MotherRecurrenceRelation}
            T_k T_\ell
            = \tfrac{1}{|W|} \sum_{w \in W} T_{k+w^\top \ell}
            = \tfrac{1}{|W|} \sum_{w \in W} T_{\ell + w^\top k},
        \end{equation}
        \item \label{item:MultivariateChebyshevSpanPolynomials} the
        multivariate Chebyshev polynomials span the space of
        multivariate polynomials,
        \item \label{item:DecompositionPropertyChebyshevPolynomials}
        the decomposition property
        \begin{equation}
            \label{eq:DecompositionPropertyChebyshevPolynomials}
            T_{k\ell \basis{e}_j} = T_{k \basis{e}_j} \circ (T_{\ell
              \basis{e}_1}, \dots, T_{\ell \basis{e}_d}),
        \end{equation}
        for $k,\ell \in \mathbb{Z}$, i.e. the generalized Chebyshev
        polynomials form a semigroup,
        \item \label{item:ChebyshevFormGroebnerBasis}
        the Chebyshev polynomials $T_{n \basis{e}_j}$, for $j = 1,
        \dots, d$, form a Gröbner basis for the ideal they generate.
    \end{enumerate}
\end{proposition}
\begin{remark}
    \label{remark:PropsOfChebyshevPolys}%
    \begin{propositionlist}
        \item \label{item:ChebyshevLieGroups} There is an intimate
        connection to differential geometry, as the permutation groups
        are a special case of so called Weyl groups associated to
        simple, simply-connected, compact Lie groups. The construction
        can indeed be carried out for every Weyl group.
        \item \label{item:ChebyshevForIntegerMatrices}
        Properties~\refitem{item:OneIsChebyshevPoly} to
        \refitem{item:MotherRecurrenceRelation} of
        Prop.~\ref{proposition:PropertiesOfMultivariateChebyshevPolynomials}
        also hold for any group of integer matrices.
        \item \label{item:ChebyshevReallyPolynomials} By looking at
        the definition it is not easy to derive that the
        multivariate Chebyshev polynomials are actually polynomials.
        This can be clarified by the recurrence relation using a
        suitable and sufficient set of starting conditions.
        \item \label{item:DecompositionProperty} The decomposition
        property
        Prop.~\ref{proposition:PropertiesOfMultivariateChebyshevPolynomials},
        \refitem{item:DecompositionPropertyChebyshevPolynomials} is
        stated in a more intricate form in
        \cite[Sect. 6]{Hoffman.Withers:1988} for Chebyshev polynomials
        associated to affine Weyl groups and in \cite[Sect.
        3]{Ricci:1986} using the algebraic generalization.  
    \end{propositionlist}
\end{remark}

We are interested in the permutation group $S_4$. This group yields
multivariate Chebyshev polynomials giving rise to a signal model on
the FCC lattice. There are $24$ possiblities to permute $4$ elements.
These permutations can be represented as $3 \times 3$ matrices, with
generator matrices
\begin{equation}
    \label{eq:GeneratorMatricesS4}%
    \begin{split}
    s_1 =
    \begin{bsmallmatrix}
        -1 & 0 & 0 \\
        1 & 1 & 0 \\
        0 & 0 & 1 
    \end{bsmallmatrix},
    s_2 =
    \begin{bsmallmatrix}
        1 & 1 & 0 \\
        0 & -1 & 0 \\
        0 & 1 & 1
    \end{bsmallmatrix},
    s_3 =
    \begin{bsmallmatrix}
        1 & 0 & 0\\
        0 & 1 & 1 \\
        0 & 0 & -1
    \end{bsmallmatrix}.        
    \end{split}
\end{equation}
These matrices satisfy
$s_i^2 = \mathbb{1}_3, (s_1 s_2)^2 = \mathbb{1}_3, (s_1 s_3)^3 =
\mathbb{1}_3, (s_2 s_3)^3 = \mathbb{1}_3$, where $\mathbb{1}_3$
denotes the $3 \times 3$ identity matrix. All other matrices,
representing a permutation of $4$ letters, can be generated by various
multiplications of these three matrices.

Let $
\begin{pmatrix}
    \theta_1 & \theta_2 & \theta_3
\end{pmatrix}
\in \mathbb{R}^3 \big/ \mathbb{Z}^3$. Using coordinates
$u := \E^{2 \pi \I \theta_1}, v := \E^{2 \pi \I \theta_2}$, and
$w := \E^ {2 \pi \I \theta_3}$, we obtain the general power form
\begin{equation*}
    \label{eq:A3PowerForm}%
    \begin{split}        
        &T_{\colvec{n}{m}{\ell}}
        = T_{n,m,\ell}(u,v,w)
        = \tfrac{1}{24} (u^n v^{l + m} w^{-l}  \\
        & + u^{-n} v^{l + m + n} w^{-l} + u^n v^m w^l \\
        & + u^{-n} v^{m + n} w^l + u^{m + n} v^l w^{-l - m} \\
        & + u^{-m - n} v^{l + m + n} w^{-l - m} + u^{l + m + n} v^{-l} 
          w^{-m} \\
        & + u^{-l - m - n} v^{m + n} w^{-m} + u^{l + m + n} v^{-l - m}
        w^m \\
        & + u^{-l - m - n} v^n w^m + u^{m + n} v^{-m} w^{l + m} \\
        & + u^{-m - n} v^n w^{l + m} + u^{l + m} v^{-l} w^{-m - n} \\
        & + u^{-l - m} v^m w^{-m - n} + u^m v^l w^{-l - m - n} \\
        & + u^{-m} v^{l + m} w^{-l - m - n} + u^l v^{-l - m} w^{-n} \\
        & + u^{-l} v^{-m} w^{-n} + u^{-l} v^{-m - n} w^n  \\
        & + u^l v^{-l - m - n} w^n + u^{l + m} v^{-l - m - n} w^{m +
          n} \\
        & + u^{-l - m} v^{-n} w^{m + n} +  u^m v^{-m - n} w^{l + m + n}
        \\
        & + u^{-m} v^{-n} w^{l + m + n})        
    \end{split}
\end{equation*}
The polynomial form can be obtained via the parametrization
\begin{align}
  x &:= \tfrac{1}{4} \left( u + u^{-1} v  + v^{-1} w + w^{-1}
      \right), \\ 
  y &:= \tfrac{1}{6} \left( v^{-1} + v + u w^{-1} + \right.\\
  \nonumber
    &\hspace{2em} \left. + u^{-1} v w^{-1}
      + u^{-1} w + u v^{-1} w \right),  \\
  z &:= \tfrac{1}{4} \left( u^{-1} + u v^{-1} + v w^{-1} + w
      \right).
\end{align}
Note that $x$ and $z$ are complex conjugates while $y$ is real since
$\overline{u} = u^{-1}, \overline{v} = v^{-1}$, and
$\overline{w} = w^{-1}$. Hence we have a \emph{real} representation if
we use the coordinates $\widetilde{x} := \tfrac{1}{2} (x + z)$ and
$\widetilde{z} := \tfrac{1}{2 \I}(x-z)$.

\begin{figure}
    \centering
     \includegraphics[width=0.35\textwidth]{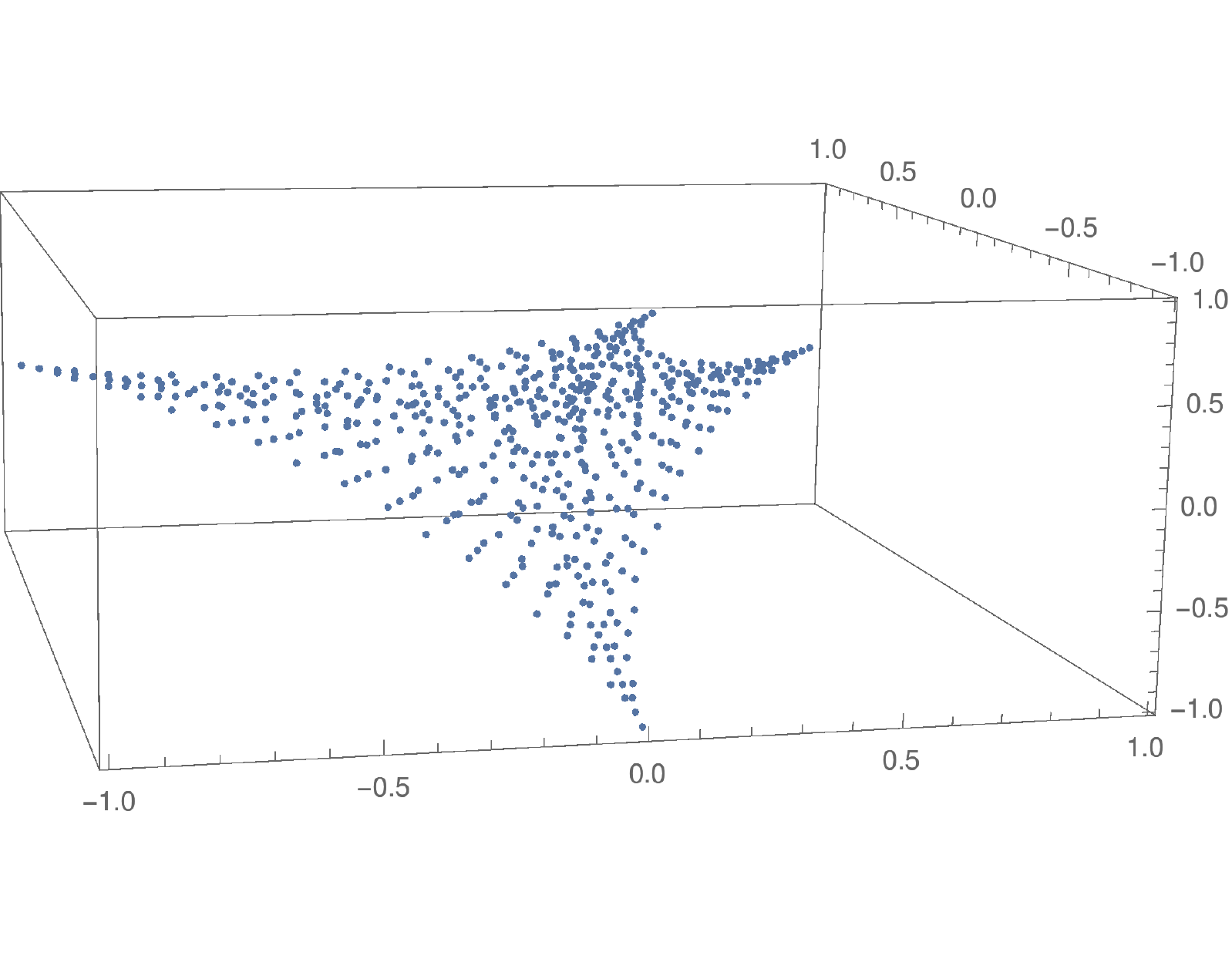}
    \caption{The common zeros of $T_{8,0,0}, T_{0,8,0}, T_{0,0,8}$.}
    \label{fig:CommonZerosA3}
\end{figure}

Using the power form we can easily deduce the common zeros of a
suitable subset of the generalized Chebyshev polynomials, which we need
for the development of a Cooley-Tukey-type-algorithm on the FCC
lattice. 
\begin{lemma}
    \label{lemma:CommonZerosA3}%
    The system of equations
    \begin{align}
      T_{n,0,0}(x,y,z) &= 0, \\
      T_{0,n,0}(x,y,z) &= 0, \\
      T_{0,0,n}(x,y,z) &= 0
    \end{align}
    has $n^3$ solutions, given in $(u,v,w)$-coordinates as
    \begin{equation}
        \label{eq:CommonZerosA3}
        (u_i, v_j, w_k) = (\omega_{8n}^{1+8i}, \omega_n^j, \omega_{8n}^{3+8k})
    \end{equation}
    with $\omega_n = \E^{2 \pi \I / n}$ being a root of unity
    and $i,j,k = 0,\dots, n-1$.
\end{lemma}
The $512$ common zeros for $n=8$ are displayed in
Fig.~\ref{fig:CommonZerosA3}, where we used the real coordinates from above.

In case of the symmetric group $S_4$ the general recurrence relation
Prop.~\ref{proposition:PropertiesOfMultivariateChebyshevPolynomials},
\refitem{item:MotherRecurrenceRelation} reads
\begin{equation*}
    \label{eq:RecurrenceRelationA3}
    \begin{split}
        & T_{j,k,p} \cdot T_{n,m,\ell}
        = \tfrac{1}{24} \left(T_{\colvec{ -j+k+n }{ -j+m+p }{ l-j }} \right.
        +T_{\colvec{  -j+k+n }{ -j+k+m-p }{ l-j }} \\ &
        +T_{\colvec{  -j+k+n }{ -j+m+p }{ l+p }}
        +T_{\colvec{  -j+k+n }{ -j+k+m-p }{ k+l-p }} 
        +T_{\colvec{  -j+k+n }{ k+m }{ l+p }} \\ &
        +T_{\colvec{  -j+k+n }{ k+m }{ k+l-p }} 
        +T_{\colvec{  -k+n+p }{ -j+m+p }{ l-j }} 
        +T_{\colvec{  n-p }{ -j+k+m-p }{ l-j }} \\ &
        +T_{\colvec{  -k+n+p }{ -j+m+p }{ l+p }} 
        +T_{\colvec{  n-p }{ -j+k+m-p }{ k+l-p }} 
        +T_{\colvec{  -k+n+p }{ m-k }{ l-j }} \\ &
        +T_{\colvec{  n-p }{ m-k }{ l-j }}
        +T_{\colvec{  -k+n+p }{ m-k }{ j-k+l }} 
        +T_{\colvec{  n-p }{ m-k }{ j-k+l }} \\ &
        +T_{\colvec{  -k+n+p }{ j-k+m+p }{ l+p }}
        +T_{\colvec{  n-p }{ j+m-p }{ k+l-p }} 
        +T_{\colvec{  -k+n+p }{ j-k+m+p }{ j-k+l }} \\ &
        +T_{\colvec{  n-p }{ j+m-p }{ j-k+l }} 
        +T_{\colvec{  j+n }{ k+m }{ l+p }} 
        +T_{\colvec{  j+n }{ k+m }{ k+l-p }} 
        +T_{\colvec{  j+n }{ j-k+m+p }{ l+p }} \\ &
        +T_{\colvec{  j+n }{ j+m-p }{ k+l-p }} 
        +T_{\colvec{  j+n }{ j-k+m+p }{ j-k+l }}
        \left. + T_{\colvec{  j+n }{ j+m-p }{ j-k+l }} \right). 
    \end{split}
\end{equation*}
This recurrence relation allows to derive the visualization of the
signal model. 

%%% Local Variables:
%%% mode: latex
%%% TeX-master: "ControloPaper"
%%% End:

%%%%%%%%%%%%%%%%%%%%%%%%%%%%%%%%%%%%%%%%%%%%%%%%%%%%%%%%%%%%%%%%%%%%%%%%%%%%%%%%
\section{FFT algorithm for FCC cosine transform}
\label{sec:CooleyTukeySU4}%

\subsection{The FCC cosine transform}
\label{subsec:SU4ChebyshevTransform}%

\begin{figure}
    \centering
    \begin{subfigure}{0.2\textwidth}        
        \includegraphics[width=\textwidth]{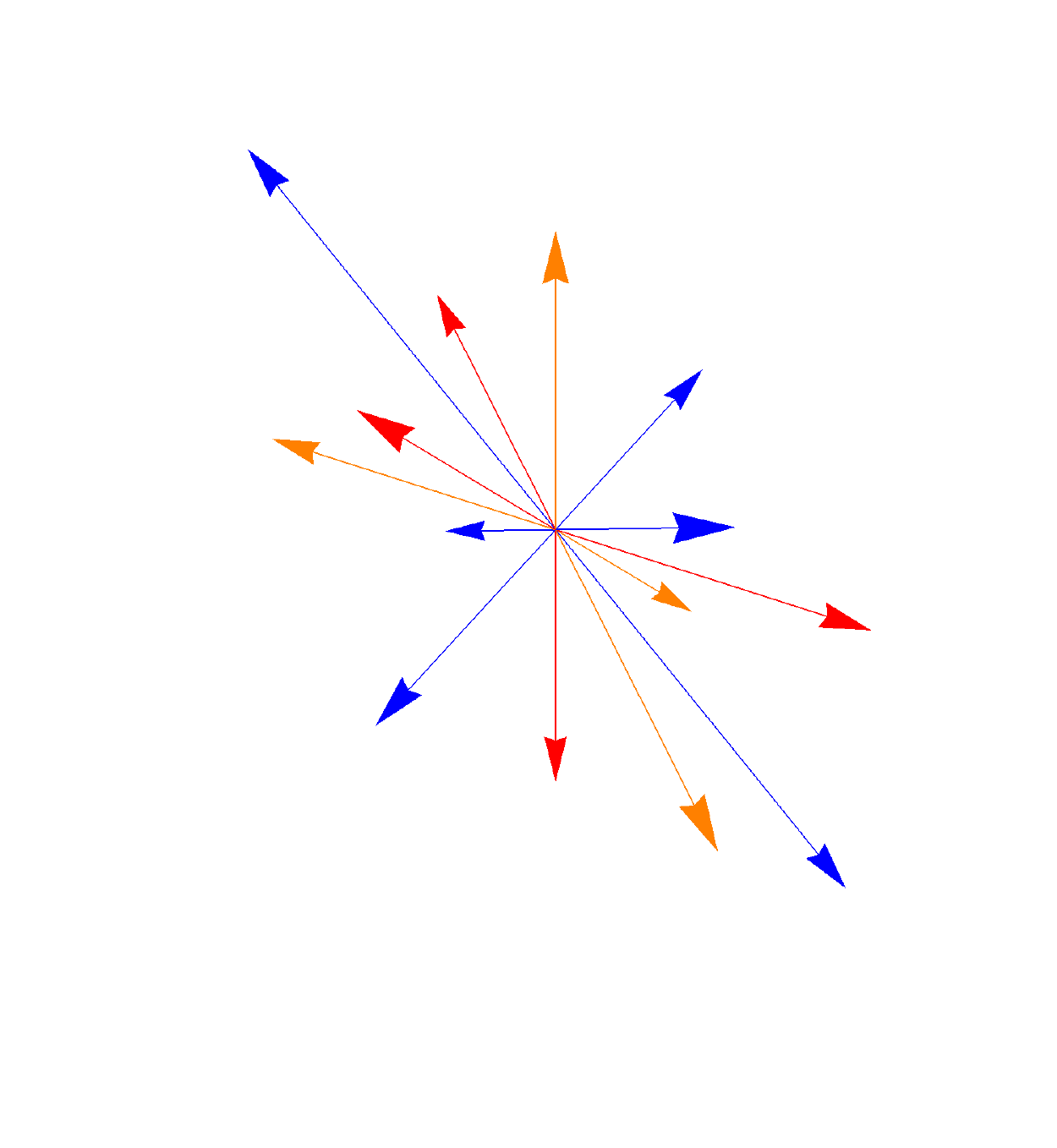}
        \caption{The shifts $T_{1,0,0}$ (red), $T_{0,1,0}$ (blue), and
          $T_{0,0,1}$ (orange).}
        \label{subfig:RootSystemA3Shifts}
    \end{subfigure}
    \hspace{0.05\textwidth}%
    \begin{subfigure}{0.2\textwidth}
        \includegraphics[width=\textwidth]{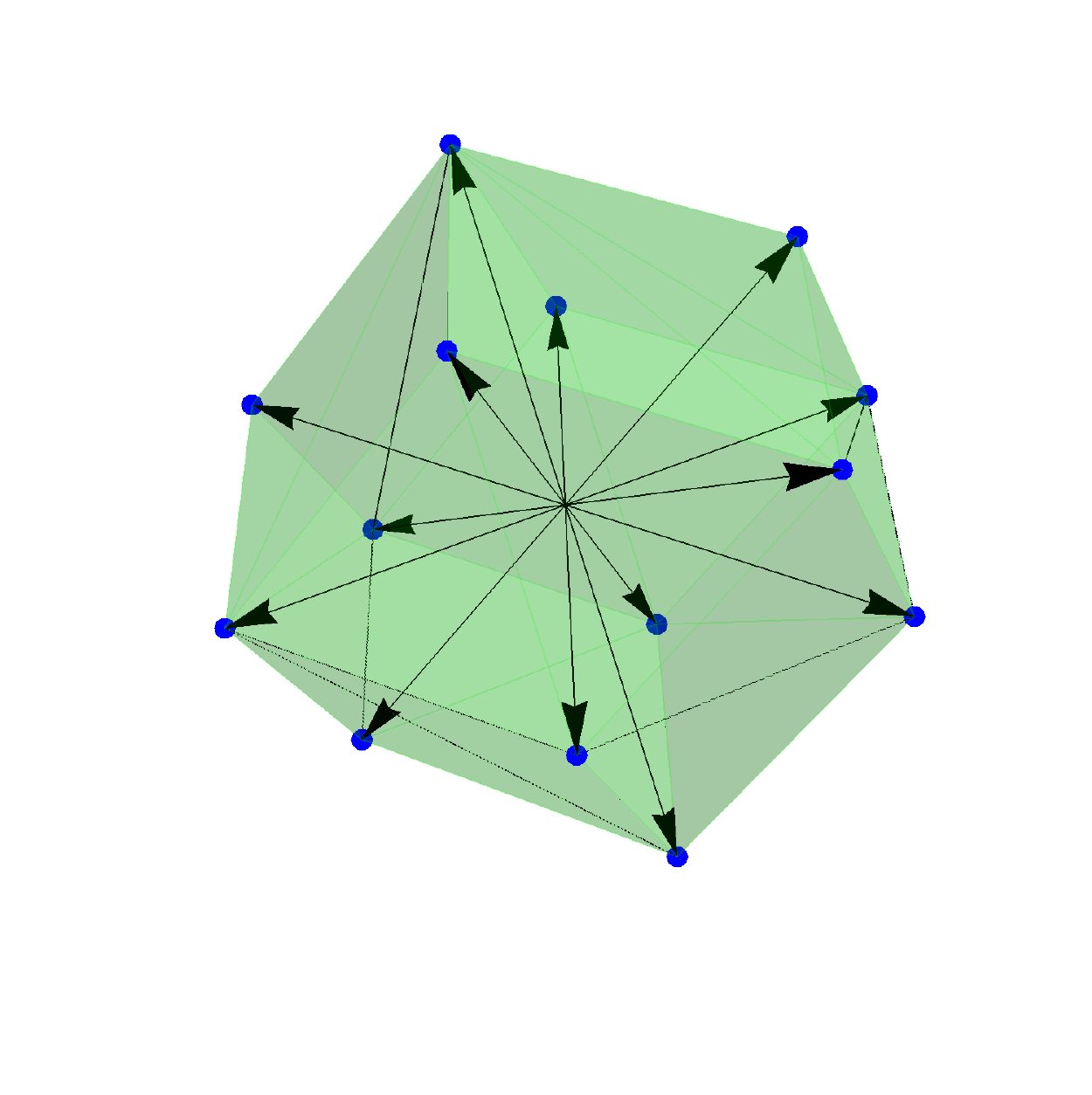}
        \caption{Convex hull of the shifts, a rhombic dodecahedron.}
        \label{subfig:ConvexHullRootSystemA3}
    \end{subfigure}
    \begin{subfigure}{0.45\textwidth}
        \includegraphics[width=\textwidth]{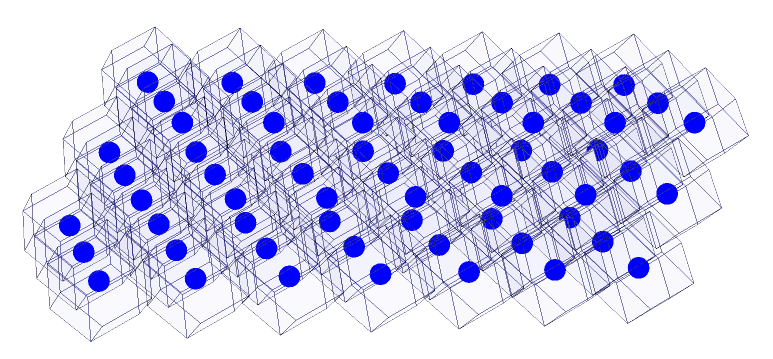}
        \caption{Space-filling via rhombic dodecahedra on FCC
          lattice points.}
        \label{subfig:ArVoronoiCells}
    \end{subfigure}
    \caption{The three variable Chebyshev polynomials give rise to
      shift operators. The rescaled convex hull of these shifts fills
      the whole space.}
    \label{fig:ShiftsAndSpaceFillingA3}
\end{figure}

We consider the signal model $M = \algebra{A} = \mathbb{C}[x,y,z]
\big/ \ideal{ T_{n,0,0}, T_{0,n,0}, T_{0,0,n}}$ with $\Phi \colon
\mathbb{C}^{n \times n \times n} \to M$ given as
\begin{equation}
    \label{eq:ZTransformFCC}
    s_{n,m,\ell} \mapsto s_{n,m,\ell} T_{n,m,\ell}.
\end{equation}
Note that $b_n := (T_{j,k,p} \; | \; 0 \leq j,k,p < n)$ is a basis of
$M$. By Lemma~\ref{lemma:CommonZerosA3} and
\eqref{eq:ChineseRemainderDecomposition} the signal space $M$
decomposes as
\begin{equation}
    \label{eq:DecompositionFCC}
    \begin{split}
        M \cong \bigoplus_{i,j,k} \mathbb{C}[x,y,z] \big/
        &\ideal{x - x(u_i,v_j,w_k), \\
        &y - y(u_i,v_j,w_k), \\
      &z - z(u_i,v_j,w_k)} .        
    \end{split}
\end{equation}
The elements $T_{1,0,0}, T_{0,1,0}$, and $T_{0,0,1}$ are the
\emph{shifts} of our signal model. If
we multiply any element of $M$ by one of these, the element
gets shifted into certain directions, as illustrated in
Fig.~\ref{subfig:RootSystemA3Shifts}. These resemble the roots of
the Lie group $SU(4)$ and generate the FCC lattice. The convex
hull of these shifts forms a rhombic dodecahedron, see
Fig.~\ref{subfig:ConvexHullRootSystemA3}. It fills the space by
rescaling and placement on the lattice points. This can be easily
seen, as the Voronoi cells of the FCC lattice are rhombic
dodecahedra, cf. \cite[Ch.~21, Sect.~3.B]{Conway.Sloane:1999}. 

We now generalize our approach by introducing \emph{skew} FCC cosine
transforms. These are necessary, as we want to develop a
divide-and-conquer approach for the algorithmic computation of the FCC
cosine transform and they appear naturally in the derivation of the
algorithm. For this, we introduce auxiliary functions
\begin{align*}
  \sigma(r,s,t) &:= \tfrac{1}{4} \left( \E^{2 \pi \I r} + \E^{2 \pi \I
                  (s-r)}  + \E^{-2 \pi \I t} + \E^{2 \pi \I
                  (t-s)}\right),  \\
  \tau(r,s,t) &:= \tfrac{1}{6} \left( \E^{-2 \pi \I s} + \E^{2 \pi \I s}
                + \E^{2 \pi \I (r-t)} + \right.  \\
  &\hspace{2em} \left. + \E^{2 \pi \I (s-r-t)} + \E^{2
                \pi \I (t-r)} + \E^{2 \pi \I (r-s+t)} \right), \\
  \rho(r,s,t) &:= \tfrac{1}{4}\left( \E^{-2\pi \I r} + \E^{2 \pi \I
                (r-s)} + \E^{2 \pi \I (s-t)} + \E^{2 \pi \I t} \right),
\end{align*}
which yield the common zeros of $T_{1,0,0}, T_{0,1,0}$, and
$T_{0,0,1}$ in $(x,y,z)$-coordinates for
$(r,s,t) = \left(\tfrac{1}{8}, 0, \frac{3}{8} \right)$, i.e. they are
all equal to zero.

Consider the filter space
$\algebra{A} = \mathbb{C}[x,y,z] \big/ \ideal{T_{n,0,0} -
  \sigma(r,s,t), T_{0,n,0} - \tau(r,s,t), T_{0,0,n} - \rho(r,s,t)}$
and the signal space $M = \algebra{A}$. If we find a decomposition of
the ideal
$\ideal{T_{n,0,0} - \sigma(r,s,t), T_{0,n,0} - \tau(r,s,t), T_{0,0,n}
  - \rho(r,s,t)}$ into coprime ideals, $M$ decomposes into
one-dimensional spaces given by the factors of the ideal by the
Chinese remainder theorem. The decomposition is achieved by finding
the common zeros of the system of equations
\begin{align}
  \nonumber
  T_{n,0,0} - \sigma(r,s,t) &= 0, \\
  \label{eq:SystemOfEquationsSkewDAT}%
  T_{0,n,0} - \tau(r,s,t) &= 0, \\
  \nonumber
  T_{0,0,n} - \rho(r,s,t) &= 0.
\end{align}
Inspecting the power form, we find the $n^3$ solutions for
\eqref{eq:SystemOfEquationsSkewDAT} in
$(u,v,w)$-parameterization as $\{ (\E^{2 \pi \I \frac{r + i}{n}},
\E^{2 \pi \I \frac{s + j}{n}}, \E^{2 \pi \I \frac{t + k}{n}}) \; | \;
i,j,k = 0, \dots, n-1 \}$. Hence we get the decomposition:
\begin{lemma}
    \label{lemma:DecompositionOfIdealA3}%
    For any $n$, we have
    \begin{equation}
        \label{eq:DecompositionOfIdealA3}%
        \begin{split}
            &\ideal[\Big]{ T_{n,0,0} - \sigma(r,s,t), T_{0,n,0} -
              \tau(r,s,t), \\
            &\quad  T_{0,0,n} - \rho(r,s,t)} \\
            &=
            \bigcap_{0 \leq i,j,k <n} \ideal[\Big]{x - \sigma\left(\tfrac{r+i}{n},
                  \tfrac{s+j}{n}, \tfrac{t+k}{n}\right), \\
              &\hspace{2em} y -
              \tau\left(\tfrac{r+i}{n}, \tfrac{s+j}{n},
                  \tfrac{t+k}{n}\right), z - \rho\left(\tfrac{r+i}{n}, 
                  \tfrac{s+j}{n}, \tfrac{t+k}{n}\right)},
        \end{split}
    \end{equation}
    and as a consequence
    \begin{equation}
      \label{eq:DecompositionOfModuleA3}%
        \begin{split}
            &\mathbb{C}[x,y,z] \Big/ \ideal[\Big]{T_{n,0,0} - \sigma(r,s,t),
              T_{0,n,0} - \tau(r,s,t), \\
              &\hspace{5em} T_{0,0,n} - \rho(r,s,t)}  \\ 
            &\cong \bigoplus_{0 \leq i,j,k < n}
            \mathbb{C}[x,y,z] \Big/ \ideal[\Big]{x -
              \sigma\left(\tfrac{r+i}{n}, \tfrac{s+j}{n},
                  \tfrac{t+k}{n}\right), \\ 
              &\hspace{2em} y - \tau\left(\tfrac{r+i}{n},
                  \tfrac{s+j}{n}, \tfrac{t+k}{n}\right), z - \rho\left(\tfrac{r+i}{n},
                  \tfrac{s+j}{n}, \tfrac{t+k}{n}\right)}. .
        \end{split}
    \end{equation} 
\end{lemma}
We now can give the definition of the FCC cosine transform:
\begin{definition}
    \label{definition:SU4ChebyshevTransform}%
    The map realizing the isomorphism
    \eqref{eq:DecompositionOfModuleA3} is called the skew FCC cosine 
    transform denoted by $\DAT_{n \times n \times n}(r,s,t)$. For
    $(r,s,t) = \left(\tfrac{1}{8}, 0, \tfrac{3}{8} \right)$, it is
    called the FCC cosine transform and is denoted by
    $\DAT_{n \times n \times n}$.
\end{definition}
An explicit description of $\DAT_{n \times n \times n}$ in matrix form
is given by inserting the common zeros of the Chebyshev polynomials
$T_{n,0,0}, T_{0,n,0}$, and $T_{0,0,n}$ into the Chebyshev polynomials
of lower order, which form a basis of $M$:
\begin{equation}
    \label{eq:SU4ChebyshevTransformExplicitForm}
    \begin{split}
        &\DAT_{n \times n \times n}(r,s,t) \\
        &= \bigg(T_{j,k,p} \left(\sigma\left(\tfrac{r+i}{n},
                \tfrac{s+\ell}{n}, \tfrac{t+q}{n}\right),
            \tau\left(\tfrac{r+i}{n},
                \tfrac{s+\ell}{n}, \tfrac{t+q}{n}\right),\right. \\
    &\hspace{2em} \left. \rho\left(\tfrac{r+i}{n},
                \tfrac{s+\ell}{n}, \tfrac{t+q}{n}\right) \right)
        \bigg)_{0 \leq i,j,k,\ell,p,q < n},
    \end{split}
\end{equation}
where $(j,k,p)$ is the row index and $(i,\ell,q)$ the column index,
both ordered lexicographically.

\subsection{The Cooley-Tukey-type-algorithm for FCC $\DAT$}
\label{subsec:CooleyTukeySU4}%

\begin{figure}
    \centering
    \includegraphics[width=0.5\textwidth]{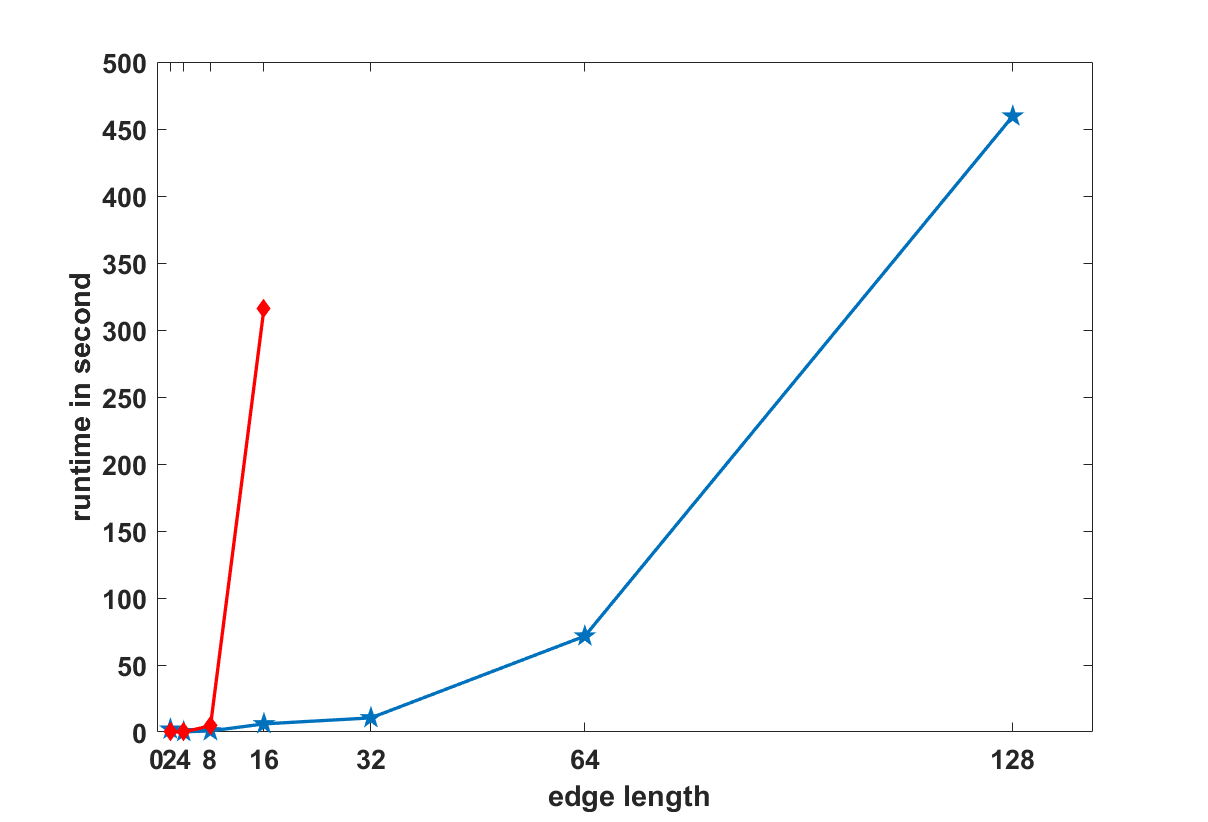}
    \caption{Runtime of a naive $O(n^6)$ implementation (red diamonds)
      and the fast $O(n^3 \log(n))$ algorithm (blue pentagrams).}
    \label{fig:NumericalExperiments}
\end{figure}

In this subsection we derive the fast radix-$2 \times 2 \times 2$
algorithm for the FCC cosine transform. Recall from
\cite[Sect.~3]{Pueschel.Roetteler:2008}, that a recursive
FFT algorithm can be deduced via stepwise decomposition
of the signal space $M$. This procedure then gives a fast algorithm if the
sample size $n$ is decomposable, e.g. $n = 2^k$ for some $k$. Applying
the stepwise composition for the case $n = 2 m$ yields
\begin{flalign}
    & \mathbb{C}[x,y,z] \big/ \\
    \nonumber
    & \ideal{T_{n,0,0} - \sigma(r,s,t), T_{0,n,0} -
      \tau(r,s,t), T_{0,0,n} - \rho(r,s,t)}   \\ 
  \label{eq:BaseChangeA3AlgebraLevel}%
  \longrightarrow &
  \mathbb{C}[x,y,z] \big/ \\
  \nonumber
  & \hspace{2em} \ideal[\big]{T_{2,0,0}(T_{m,0,0}, T_{0,m,0}, T_{0,0,m}) -
    \sigma(r,s,t), 
     \\
  \nonumber
  & \hspace{2em} T_{0,2,0}(T_{m,0,0}, T_{0,m,0}, T_{0,0,m}) - \tau(r,s,t),  \\
  \nonumber
  & \hspace{2em} T_{0,0,2}(T_{m,0,0}, T_{0,m,0}, T_{0,0,m}) - \rho(r,s,t)}   \\
  \label{eq:BaseCaseA3AlgebraLevel}%
  \longrightarrow &
  \bigoplus_{i_r, i_s, i_t = 0,1} \mathbb{C}[x,y,z]
  \big/ \\
  \nonumber
  &\hspace{2em}
  \ideal[\Big]{T_{m,0,0} - \sigma\left( \tfrac{r + i_r}{2}, \tfrac{s +
  i_s}{2}, \tfrac{t + i_t}{2} \right),  \\
  \nonumber
  & \hspace{2em} T_{0,m,0} - \tau\left( \tfrac{r + i_r}{2}, \tfrac{s +
        i_s}{2}, \tfrac{t + i_t}{2} \right),  \\
  \nonumber
  & \hspace{2em} T_{0,0,m} - \rho\left( \tfrac{r + i_r}{2}, \tfrac{s +
        i_s}{2}, \tfrac{t + i_t}{2} \right) }  \\
  \label{eq:RecursionStepA3AlgebraLevel}%
  \longrightarrow &
  \bigoplus_{i_r, i_s, i_t = 0,1 \,} \bigoplus_{\, k_r, k_s, k_t = 0,\dots,
    m-1}
  \mathbb{C}[x,y,z] \big/ \\
  \nonumber
  & \hspace{2em} \ideal[\Big]{ x - \sigma\left( \tfrac{r + i_r + 2
  k_r}{n}, \tfrac{s + i_s + 2 k_s}{n}, \tfrac{t + i_t + 2 k_t}{n}
  \right),  \\
  \nonumber
  & \hspace{2em} y - \tau\left( \tfrac{r + i_r + 2 k_r}{n}, \tfrac{s +
        i_s + 2 k_s}{n}, 
  \tfrac{t + i_t + 2 k_t}{n} \right),  \\
  \nonumber
  & \hspace{2em} z - \rho\left( \tfrac{r + i_r + 2 k_r}{n}, \tfrac{s + i_s + 2
  k_s}{n}, \tfrac{t + i_t + 2 k_t}{n} \right) }  \\
  \label{eq:PermutationStepA3AlgebraLevel}%
  \longrightarrow &
  \bigoplus_{j_r, j_s, j_t = 0, \dots, n-1}
  \mathbb{C}[x,y,z] \big/  \\
  \nonumber
  & \hspace{2em} \ideal[\Big]{ x - \sigma\left( \tfrac{r + j_r}{n},
        \tfrac{s + j_s}{n}, \tfrac{t + j_t}{n} \right),  \\
  \nonumber
  & \hspace{2em} y - \tau\left( \tfrac{r + j_r}{n}, \tfrac{s + j_s}{n},
      \tfrac{t + j_t}{n} \right), %  \\
  % \nonumber
  % & \hspace{2em}
  z - \rho\left( \tfrac{r + j_r}{n}, \tfrac{s + j_s}{n},
      \tfrac{t + j_t}{n} \right) }.
\end{flalign}
Each step is encoded by a matrix. We have for
\eqref{eq:BaseChangeA3AlgebraLevel} a complicated basis change matrix
$B_n(r,s,t)$. We go from the old basis $b_n$ ordered
lexicographically to the new basis
\begin{equation*}
    \widetilde{b}_n :=
    \begin{bsmallmatrix}
        T_{0,0,0} T_{0,0,0}(T_{m,0,0}, T_{0,m,0}, T_{0,0,m})
        \\
        \vdots  \\
        T_{m-1,m-1,m-1} T_{1,1,1}(T_{m,0,0}, T_{0,m,0}, T_{0,0,m})
    \end{bsmallmatrix}.
\end{equation*}
This basis change, which has $O(n^3)$ entries, will be described in
future works. For \eqref{eq:BaseCaseA3AlgebraLevel} we apply the
matrix $\DAT_{2 \times 2 \times 2}(r,s,t) \tensor \mathbb{1}_{m^3}$,
for \eqref{eq:RecursionStepA3AlgebraLevel} we obtain the recursion
steps via the application of
$\bigoplus_{i_r, i_s, i_t = 0,1} \DAT_{m \times m \times m}\left(
    \frac{r + i_r}{2}, \frac{s + i_s}{2}, \frac{t + i_t}{2} \right)$.
Finally we get a permutation matrix $P_n$ in
\eqref{eq:PermutationStepA3AlgebraLevel}. The fast algorithm is
given as the matrix factorization
\begin{equation}
    \label{eq:FastAlgorithm}
    \begin{split}
        &\DAT_{n \times n \times n}(r,s,t) \\
        &= P_n
        \cdot \bigoplus_{i_r, i_s, i_t = 0,1} \DAT_{m \times m \times m}\left(
            \tfrac{r + i_r}{2}, \tfrac{s + i_s}{2}, \tfrac{t + i_t}{2}
        \right) \\        
        &\quad \cdot (\DAT_{2 \times 2 \times 2}(r,s,t) \tensor
        \mathbb{1}_{m^3}) 
        \cdot B_n(r,s,t).        
    \end{split}
\end{equation}
Looking at the matrix representation makes it easy to see that the
whole algorithm has cost $O(n^3 \log(n))$. Hence it can compete with
the tensor product of standard Cooley-Tukey FFTs or fast Cosine
transforms. First numerical experiments verify the effectivity of the
proposed methodology compared to the naive $O(n^6)$ implementation, as
depicted in Fig.~\ref{fig:NumericalExperiments}.

%%% Local Variables:
%%% mode: latex
%%% TeX-master: "ControloPaper"
%%% End:

%%%%%%%%%%%%%%%%%%%%%%%%%%%%%%%%%%%%%%%%%%%%%%%%%%%%%%%%%%%%%%%%%%%%%%%%%%%%%%%%
%\section{Numerical tests}
%\label{sec:Numerics}%

%\input{VoxelApplication}

%%%%%%%%%%%%%%%%%%%%%%%%%%%%%%%%%%%%%%%%%%%%%%%%%%%%%%%%%%%%%%%%%%%%%%%%%%%%%%%%
\section{Application to voxel data and future work}
\label{sec:Application}%

One of the main advantages of using the algebraic signal model is
that we do not actually have to sample data on the FCC lattice, but
can instead use the $z$-transform to map any data on it. Hence we can
compare the effect of the FCC transform directly with the standard
approach to discrete cosine transforms on cc lattices.

\begin{figure}
    \centering
    \begin{subfigure}{0.43\textwidth}
        \includegraphics[width=0.9\textwidth]{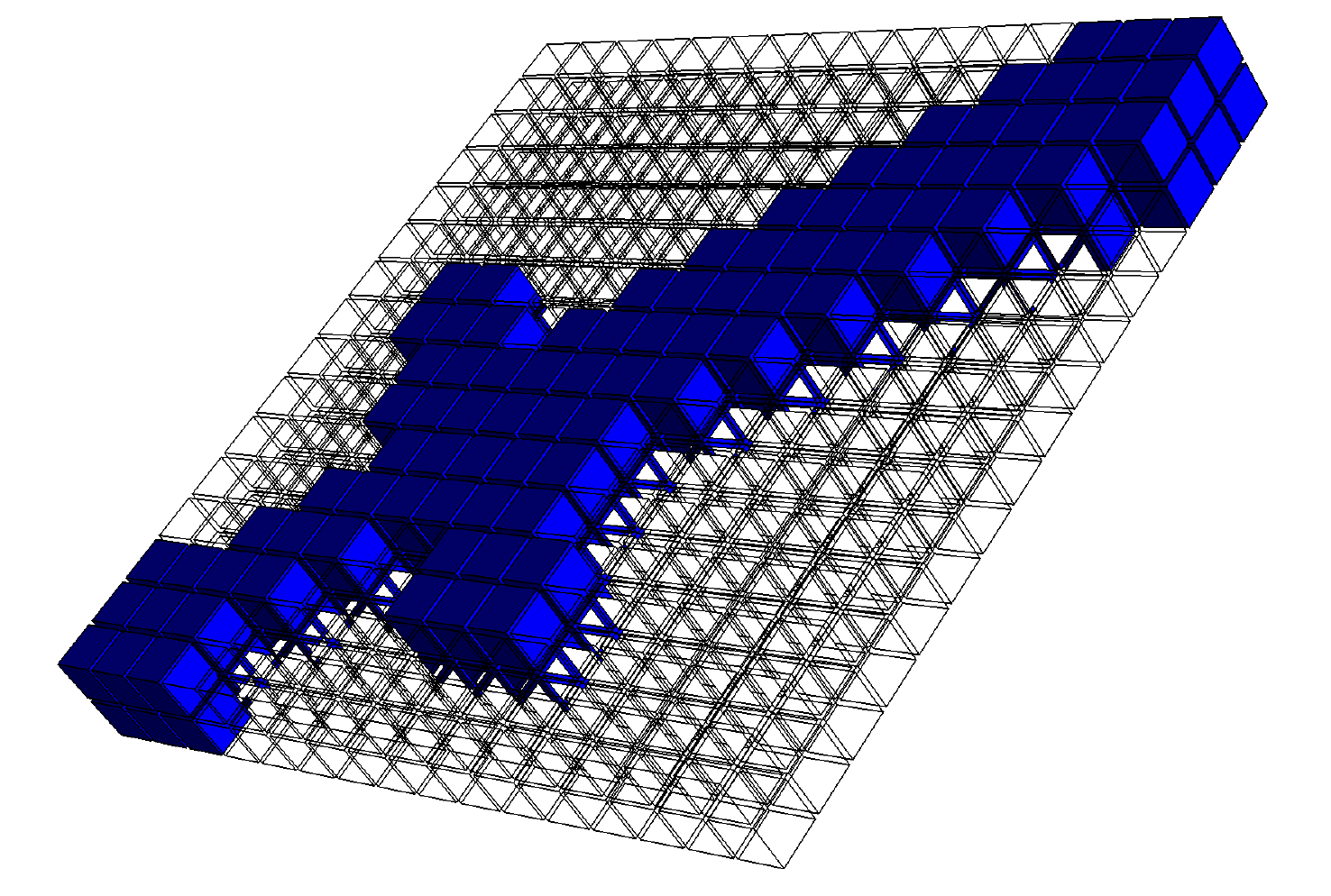}
        \vspace{0.5em}
        \caption{Sword represented via cuboids.}
        \label{subfig:SwordRect}
    \end{subfigure}
    \begin{subfigure}{0.43\textwidth}
        \includegraphics[width=1.2\textwidth]{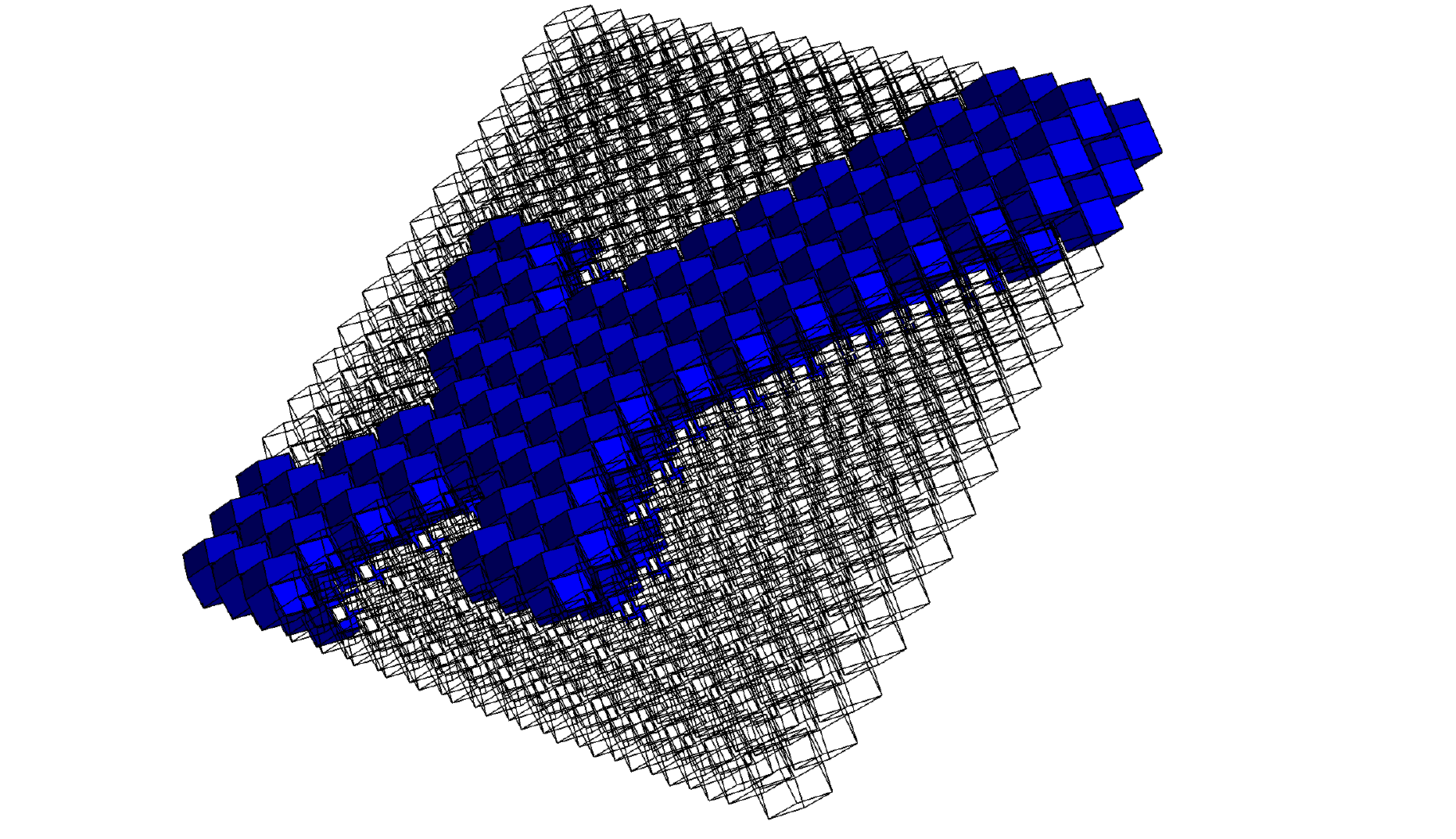}
        \caption{Sword represented via rhombic dodecahedra.} 
        \label{subfig:SwordA3}
    \end{subfigure}
    \caption{Data representation on a rectangular grid
       and on a $SU(4)$ grid.} 
    \label{fig:MinecraftSword}
\end{figure}

\begin{figure}
    \centering
    \begin{subfigure}{0.43\textwidth}
        \includegraphics[width=\textwidth]{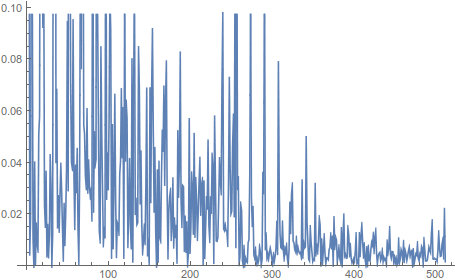}
        \vspace{0.5em}
        \caption{The threefold discrete cosine spectrum of the sword on a
          CC lattice.}
        \label{subfig:DCTSpectrumSword}
    \end{subfigure}
    \begin{subfigure}{0.43\textwidth}
        \includegraphics[width=\textwidth]{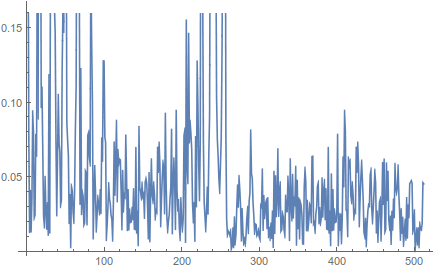}
        \caption{The $\DAT$ spectrum of the sword on a FCC lattice.} 
        \label{subfig:DATSpectrumSword}
    \end{subfigure}
    \caption{Spectra of the Sword.}
    \label{fig:SwordSpectra}
\end{figure}

For graphic representation of voxel data on a rectangular grid, each
data point gets assigned to a cuboid and can define certain
properties, like color or opacity. Now for the FCC lattice we have
to assign the points to rhombic dodecahedra instead, as they fill the
space on this lattice. The difference in the graphical representation
is shown in Fig~\ref{fig:MinecraftSword}, where we created an
artifical data set portraying a sword. The data values were mapped to
opacity values, i.e. data points with value $0$ are not visible, while
the ones with $1$ are fully visible.
In Fig.~\ref{subfig:DCTSpectrumSword} we see the effect of the
threefold tensor product of the discrete cosine transform and in
Fig.~\ref{subfig:DATSpectrumSword} the effect of the FCC Chebyshev
transform.
Aside from an explicit description of the matrices used in the
decomposition steps, we plan to derive Cooley-Tukey-type-algorithms
based on generalized Chebyshev polynomials for all simple, compact Lie
groups.

%%% Local Variables:
%%% mode: latex
%%% TeX-master: "FastAdLattice"
%%% End:

%\newpage
%%%%%%%%%%%%%%%%%%%%%%%%%%%%%%%%%%%%%%%%%%%%%%%%%%%%%%%%%%%%%%%%%%%%%%%%%%%%%%%%
%\nocite{*}
\bibliographystyle{IEEEtran}
\bibliography{Controlo}

%%%%%%%%%%%%%%%%%%%%%%%%%%%%%%%%%%%%%%%%%%%%%%%%%%%%%%%%%%%%%%%%%%%%%%%%%%%%%%%%

\addtolength{\textheight}{-12cm}   % This command serves to balance
% the column lengths

\end{document}